\documentclass[10pt,aps,twocolumn,superscriptaddress,prl]{revtex4}

\usepackage{graphicx,epsfig}
\usepackage{amssymb}
\begin{document}
\title{Quasiparticle Tunneling in the Fractional Quantum Hall State at $\nu=5/2$}

\author{Iuliana P. Radu}
\affiliation{Department of Physics, Massachusetts Institute of
Technology, Cambridge, Massachusetts 02139}
\author{J. B. Miller}
\affiliation{Department of Physics, Harvard University, Cambridge,
Massachusetts 02138}
\author{C. M. Marcus}
\affiliation{Department of Physics, Harvard University, Cambridge,
Massachusetts 02138}
\author{M. A. Kastner}
\affiliation{Department of Physics, Massachusetts Institute of
Technology, Cambridge, Massachusetts 02139}
\author{L. N. Pfeiffer}
\affiliation{Bell Labs, Lucent Technologies, Murray Hill, New Jersey
07974}
\author{K. W. West}
\affiliation{Bell Labs, Lucent Technologies, Murray Hill, New Jersey
07974}

\date{7 March 2008}

\begin{abstract}

Theory predicts that quasiparticle tunneling between the
counter-propagating edges in a fractional quantum Hall state can be
used to measure the effective quasiparticle charge $e^*$ and
dimensionless interaction parameter $g$, and thereby characterize
the many-body wavefunction describing the state.  We report
measurements of quasiparticle tunneling  in a high mobility GaAs
two-dimensional electron system in the fractional quantum Hall state
at $\nu=5/2$ using a gate-defined constriction to bring the edges
close together.  We find the dc-bias peaks in the tunneling
conductance at different temperatures collapse onto a single curve
when scaled, in agreement with weak tunneling theory. Various models
for the $\nu=5/2$ state predict different values for $g$. Among
these models, the non-abelian states with $e^*=1/4$ and $g=1/2$ are
most consistent with the data.
\end{abstract}


\maketitle
\noindent

The fractional quantum Hall (FQH) effect \cite{FQHE} results from
the formation of novel states of a two-dimensional electron system
(2DES) at high magnetic field and low temperature, in which
electron-electron interactions lead to gaps in the bulk excitation
spectra. Because of these gaps, current can only flow via extended
states that propagate around the edges of the 2DES
\cite{Halperin_edges}. At a constriction in the 2DES, such as that
formed by a quantum point contact (QPC), counter-propagating edge
states come close enough together that quasiparticles can tunnel
between them. According to theory \cite{wenPRB44:1991}, weak
quasiparticle tunneling depends strongly on the voltage difference
between the edges (or, because of the Hall effect, the current
through the QPC), and should scale with temperature in a way that
provides a measurement of the effective charge, $e^*$, of the
quasiparticles and the strength of the Coulomb interaction, $g$.
Since both $e^*$ and $g$ are specific to the FQH state, such
measurements provide a discriminating probe of FQH wavefunctions.

\begin{figure}[b]
\includegraphics[width=3.25in]{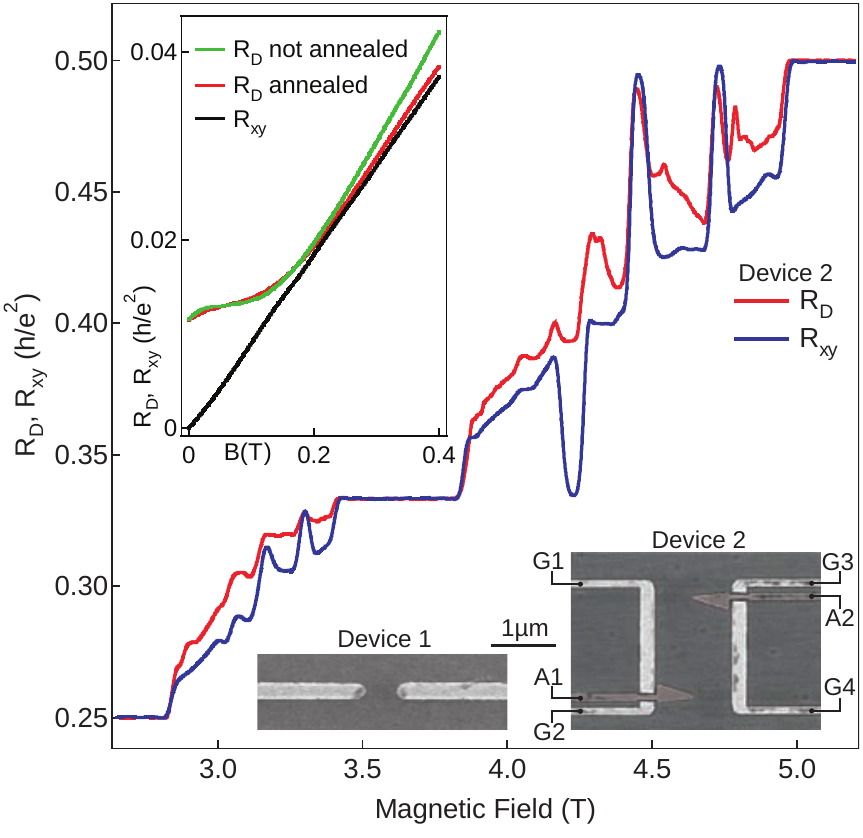}
\caption{
Magnetic field dependence of the diagonal ($R_D$) and Hall ($R_{xy}$) resistance for device 2 at fixed gate voltage from $\nu=2$ to $\nu=4$ illustrating that both the QPC and the bulk are at the same filling fraction.  The upper inset shows low-field data from the same device (device 2) emphasizing that the carrier density in the annealed QPC is nearly the same as that of the bulk (red and black traces with almost matching slopes), while in the non-annealed QPC (green trace) the density shifts significantly. For clarity, the non-annealed data has been offset vertically by $0.003$~h/e$^2$. Lower insets are scanning electron micrographs of devices with similar gate geometry to those used in these experiments. In device 2, grounded gates held are artificially colored gray.}
\end{figure}

\begin{figure*}
\includegraphics[width=12cm]{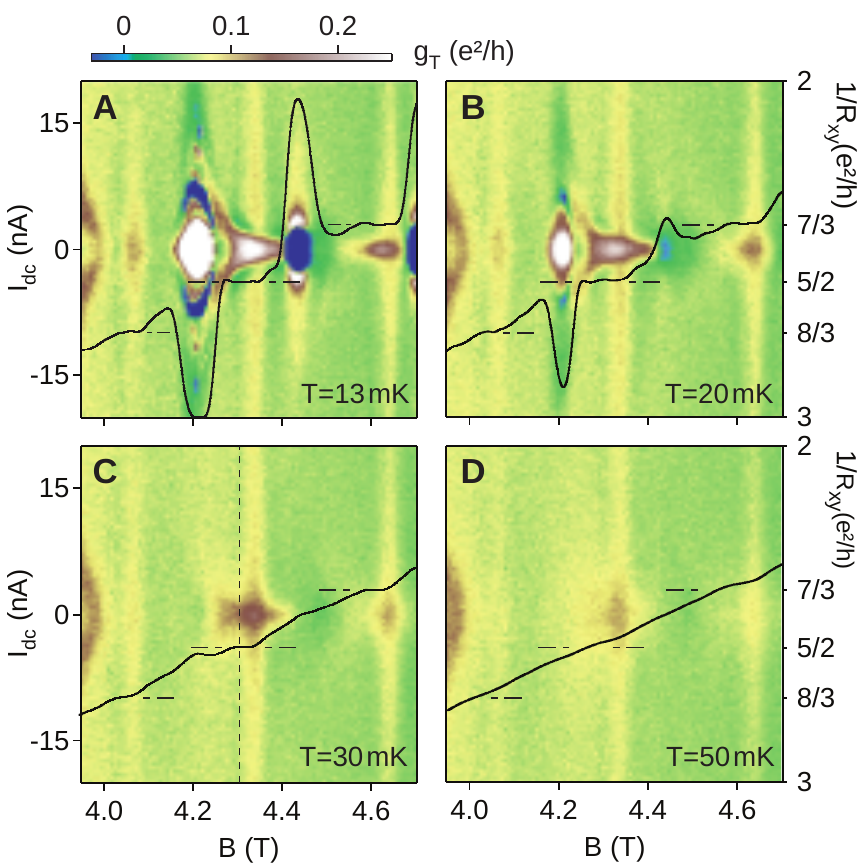}
\caption{Differential tunneling conductance $g_T$ (device 2) as a function of magnetic field and dc bias current at several temperatures (marked on figure).  On each graph, the zero-dc-bias $R_{xy}$ trace from the same temperature is superimposed (right axis). The field range encompasses the FQH states $7/3$, $5/2$ and $8/3$ (marked with horizontal dot-dash lines). At the higher temperatures, dc bias non-linearities exist only at the fractional plateaus. All other features, such as those from the re-entrant quantum Hall effect, disappear at $\sim$ 30~mK. }
\end{figure*}

The FQH state at $\nu=5/2$ \cite{Willett1987} has received
particular attention because the leading candidates for the
wavefunction for this state have elementary excitations that exhibit
non-abelian particle statistics \cite{Haldane_rezayi, moore_read,
PhysRevLett.66.802, levin:236806, lee:236807}. Whereas the
interchange of abelian particles such as electrons multiplies the
wavefunction by an overall phase, the interchange of non-abelian
quasiparticles can lead to a different wavefunction.  Identifying a
physical system with non-abelian statistics would be of fundamental
interest, but would also provide a basis for a topological quantum
information processing scheme \cite{Kitaev9707021} that is resistant
to environmental decoherence \cite{dassarma2005, revmodphys}.
Although wavefunctions with non-abelian excitations are the prime
candidates \cite{Morf} to describe the state at $\nu=5/2$,
alternatives with abelian properties have also been proposed
\cite{Halperin_orig331, Halperin331, K8}.  All candidate
wavefunctions for $\nu=5/2$ have quasiparticle effective charge
$e^*=1/4$, but they differ in the predicted values of $g$
\cite{PhysRevLett.70.355,bishara2,fendley2006,levin:236806,lee:236807}.

Weak tunneling theory, developed originally for Laughlin FQH states \cite{wenPRB44:1991}, has also been extended to non-abelian states \cite{read, wen:advances, PhysRevLett.70.355, bishara2, fendley2006}. Tunneling measurements on a single constriction can distinguish among candidate wavefunctions for $\nu=5/2$; existing proposals to find direct evidence for non-abelian statistics, however, require multiple constrictions to create interference among tunneling paths  \cite{fradkin, dassarma2005, bonderson2006, stern:016802, hou:146802, feldman2006}.

Experimentally, the quasiparticle charge, $e^*$, has been investigated for FQH states at $\nu<1$ using shot noise \cite{Heiblum1, French} and interferometry \cite{Su}, yielding results generally consistent with theory.  Recently, a measurement of quasiparticle charge for the $\nu = 5/2$ state, also using shot noise, have obtained values consistent with $e^* = 1/4$  \cite{merav}. Previous experiments of quasiparticle tunneling at a constriction have focused on cases of unequal filling fractions in the bulk and the constriction \cite{Roddaro1, Roddaro2, Roddaro3, Jeff}. These experiments identify zero-bias features associated with quasiparticle tunneling at FQH edges, and are compared to the present measurements below. Finally, the interaction parameter, $g$, has been measured in studies of tunneling of $\nu=1/3$ FQH edges through a depleted constriction \cite{webb} through which electrons, rather than quasiparticles, tunnel.

In this article, we present experimental measurements of quasiparticle tunneling at a QPC at $\nu=5/2$, in the regime where the filling fraction (and carrier density) in the QPC and the bulk 2DES are the same. We find that  tunneling conductance across the QPC exhibits a strong zero-bias peak that scales with temperature in quantitative agreement with the theory for weak tunneling \cite{wenPRB44:1991, bishara2, fendley2006}. From these measurements, we extract $e^*$ and $g$. We observe that among the candidate states for  $\nu=5/2$,  the anti-Pfaffian \cite{lee:236807, levin:236806} and the $U(1)\times SU_2(2)$ \cite{PhysRevLett.66.802} are most consistent with the data.

\paragraph*{Sample and experimental setup.} The sample is a GaAs/AlGaAs heterostructure with the 2DES 200~nm below the surface and two Si $\delta$-doping layers 100~nm above and below the 2DES. Hall bars with a width of 150~$\mu$m are patterned on this heterostructure. The mobility (before the gates are energized) is 2000~m$^{2}$/Vs, the carrier density is 2.6$\times$10$^{15}$~m$^{-2}$, and the $\nu=5/2$ energy gap is $\sim$130~mK in the bulk \cite{Jeff}. The QPCs are formed by Cr/Au top gates, which are patterned on the Hall bar using e-beam lithography. By applying a negative gate voltage $V_g$ to these gates, the electrons underneath them are depleted, creating a constriction tunable with $V_g$. We report measurements on devices with two different gate geometries (lower insets of Fig. 1). Device 1 is a simple QPC with gate separation of 800~nm. Device 2 is a channel  $\sim$ 1200~nm wide, formed by energizing the gates marked G1, G2, G3 and G4 (gates A1 and A2 are held at !
 ground and not used in this experiment). The sample is mounted on the cold finger of a dilution refrigerator with a base temperature of less than 10~mK. In all figures and analysis, we quote electron temperatures. At temperatures $\geq20$~mK, the mixing chamber and electron temperatures have been measured to be equal using resonant electron tunneling in a lateral quantum dot. Temperatures below 20~mK have been estimated using both resonant tunneling and by tracking several strongly temperature-dependent quantum Hall features in the bulk, with consistent results. (See Supporting Online Material.) The magnetic field is oriented perpendicular to the plane of the 2DES.

Measurements are performed using standard 4-probe lock-in techniques with an ac current excitation between 100 - 400~pA and in some cases a dc bias current of up to 20~nA. To determine the tunneling conductance $g_T$, we simultaneously measure the Hall resistance $R_{xy}$ (voltage probes on opposite sides of the Hall bar away from the QPC) and the diagonal resistance $R_{D}$ (voltage probes on opposite sides of the Hall bar and also opposite sides of the QPC)\cite{Beenakker, Jeff}. In the weak tunneling regime \cite{wenPRB44:1991}, when the bulk of the sample is at a quantum Hall plateau, the tunneling voltage is the same as the Hall voltage, while $R_D$ reflects the differential tunneling conductance via: \begin{equation}\label{}
    g_T = \frac{R_D-R_{xy}}{R_{xy}^2}
\end{equation}
\noindent Note that $R_{xy}$ is independent of dc bias when the bulk is at a FQH plateau. If one assumes that the underlying edge has a filling fraction $\nu_{\mathrm{under}}$, then the reflection of the $5/2$ edge state can be calculated as: $\mathcal{R}= g_TR_{xy}^2/[(1/\nu_{\mathrm{under}})h/e^2 - R_{xy}]$.

\paragraph*{Same filling fraction in QPC and bulk.} A key difference between this work and previous tunneling experiments \cite{Roddaro1, Roddaro2, Roddaro3, Jeff} is that we are able to deplete the electrons under the gates and induce tunneling without significantly changing the filling fraction in the QPC. This is achieved by applying a gate voltage of -3~V while at 4~K and allowing the system to relax for several hours, which we refer to as annealing.  We then cool the sample and limit the voltage to the range -2 to -3~V when at dilution refrigerator temperatures. After annealing, $R_D$ and $R_{xy}$ are measured over several integer plateaus and the fields marking the ends of the plateaus are found to coincide for QPC and the bulk (see Fig.~1), indicating that the filling factors are the same. The extra resistance in $R_D$ at FQH states is consistent with tunneling, as discussed below. Additional evidence that the filling fraction changes little once the QPC is annealed i!
 s shown in the inset of Fig.~1:  the slopes of $R_{xy}$ and $R_D$ at low magnetic field, inversely proportional to carrier density, differ by 2\% or less. For comparison, we show data from a non-annealed QPC where the density decreases by $\sim$15\%.

\begin{figure}
\includegraphics[width=3.25in]{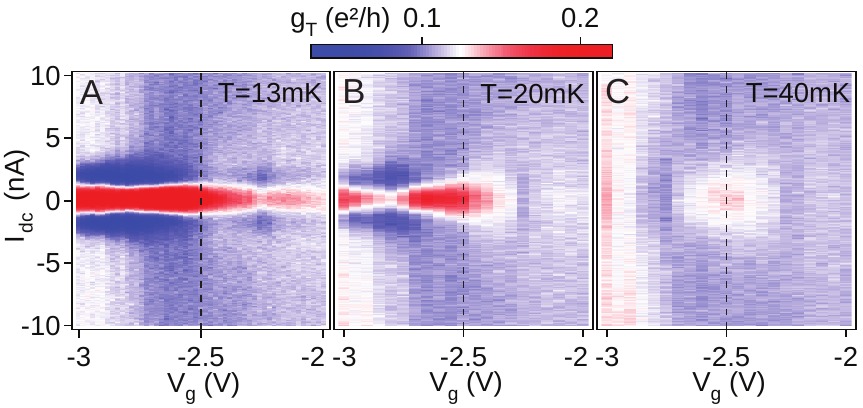}
\caption{
Differential tunneling conductance $g_T$ (device 1) as a function of $V_g$ and dc bias current at several temperatures: {\bf A.} $T=13$~mK, {\bf B.} $T=20$~mK, {\bf C.} $T=40$~mK. A peak in both dc bias and $V_g$ becomes visible at T=40~mK. The vertical dashed line marks the center of this resonance. }
\end{figure}

\paragraph*{Bias and temperature dependence}
In the following, we focus on the dependence of $R_D$ on the dc current bias $I_{dc}$ through the QPC and Hall bar. Figure~2 shows a color-scale plot  of the dependence of $R_D$ on both $I_{dc}$ and magnetic field $B$ at four temperatures; a measurement of $R_{xy}$ is shown for comparison. As seen most clearly at the highest temperatures, these field sweeps reveal a series of FQH states \cite{xia2004} around $\nu=5/2$, including the 7/3 and 8/3.  At the lowest temperatures strong re-entrant integer quantum Hall (RIQH) features are also visible on either side of 5/2, both in the bulk and in the QPC (see Fig. 2). The dc bias behavior at FQH plateaus is quite different from that of the RIQH features: At FQH plateaus, zero-bias peaks in $g_T$ persist up to at least 50~mK (Fig.~2D). By contrast, RIQH states have more complex dc bias signatures, which decrease rapidly with temperature, disappearing by 30~mK both in the bulk ($R_{xy}$) and in the QPC ($g_T$). Qualitatively similar !
 results are observed for device 1. To study the FQH state at $\nu=5/2$, we set the magnetic field to the center of a bulk FQH plateau ($B=4.31$~T for device 2, vertical line in Fig.~2C, and $B=4.3$~T for device 1).

\begin{figure*}
\includegraphics[width=11cm]{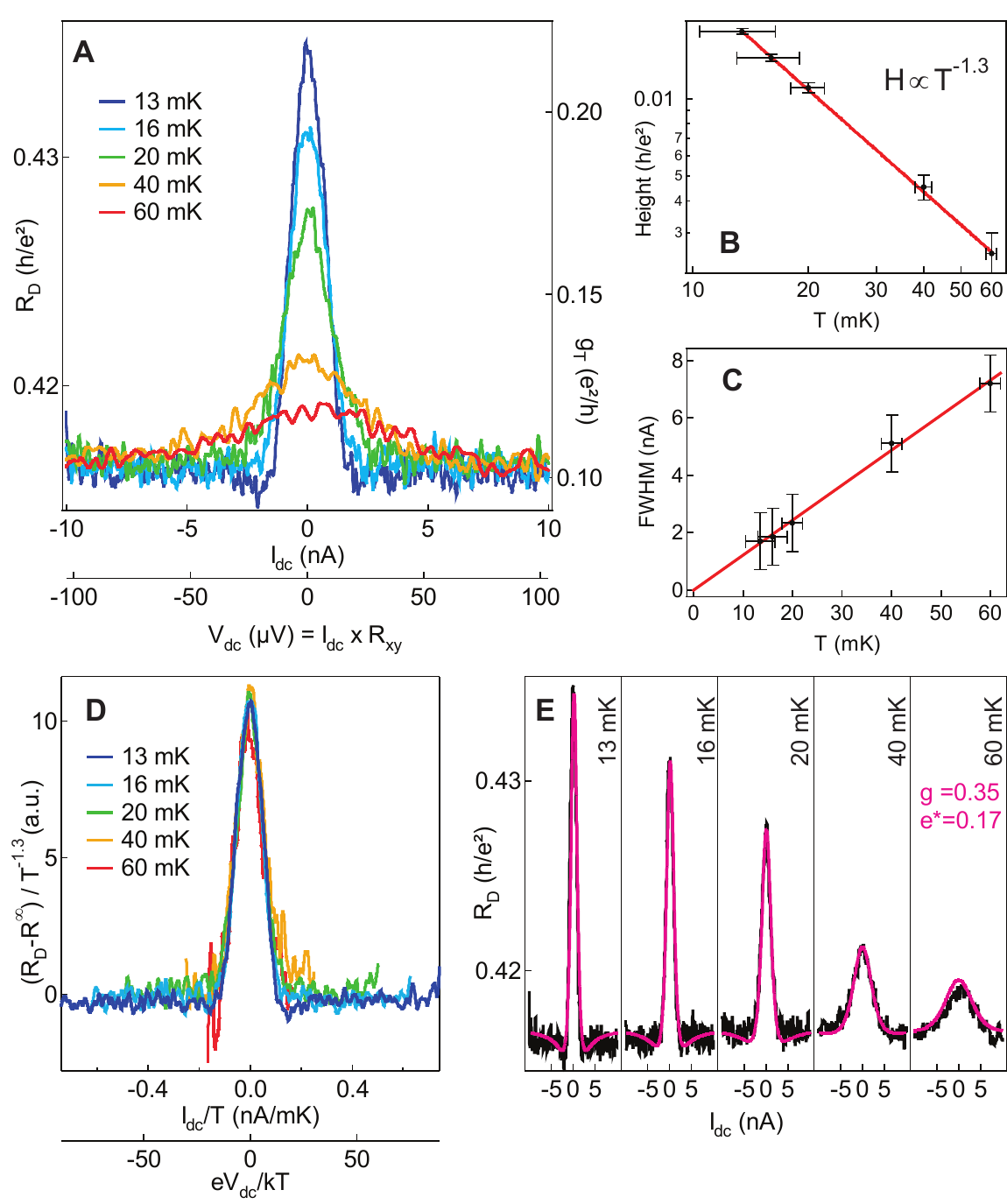}
\caption{
{\bf A.} $R_{D}$ (device 1) as a function of dc bias at fixed magnetic field ($B=4.3$~T, middle of $\nu=5/2$) and fixed gate voltage ($V_g=-2.5$~V) at several temperatures. $R_{xy}$ is independent of dc bias over the range of $I_{dc}$ (not shown), which makes the bias dependence of $R_{D}$ proportional to that of $g_T$ (right axis) up to a constant. {\bf B.} Zero dc bias peak height as a function of temperature. The red line is the best fit with a power law where the exponent is -1.3. {\bf{C.}} The peak full width at half maximum (FWHM) as a function of temperature. The red line is the best fit with a line going through zero. {\bf{D.}} Data collapsed onto a single curve using an exponent -1.3. {\bf{E.}} Best fit of all the data in {\bf{A}} with the weak tunneling formula (eq. 2) returns $e^*=0.17$ and $g=0.35$.}
\end{figure*}

With the field set to the center of the plateau, the effect of $V_g$ on the zero-bias peak at several temperatures is presented in Fig.~3. At the lowest temperatures (Fig.~3A), the zero-bias peak persists throughout the $V_g$ range.  At higher temperatures, a peak in both dc bias and $V_g$ is observed, centered near $V_g=-2.5$~V (Fig.~3C). To study quasiparticle tunneling, we set $V_g$ to the center of this peak, the feature that persists to the highest temperature, because theory predicts that tunneling decreases slowly, as a power law, with temperature.

\begin{figure}
\includegraphics[width=3.25in]{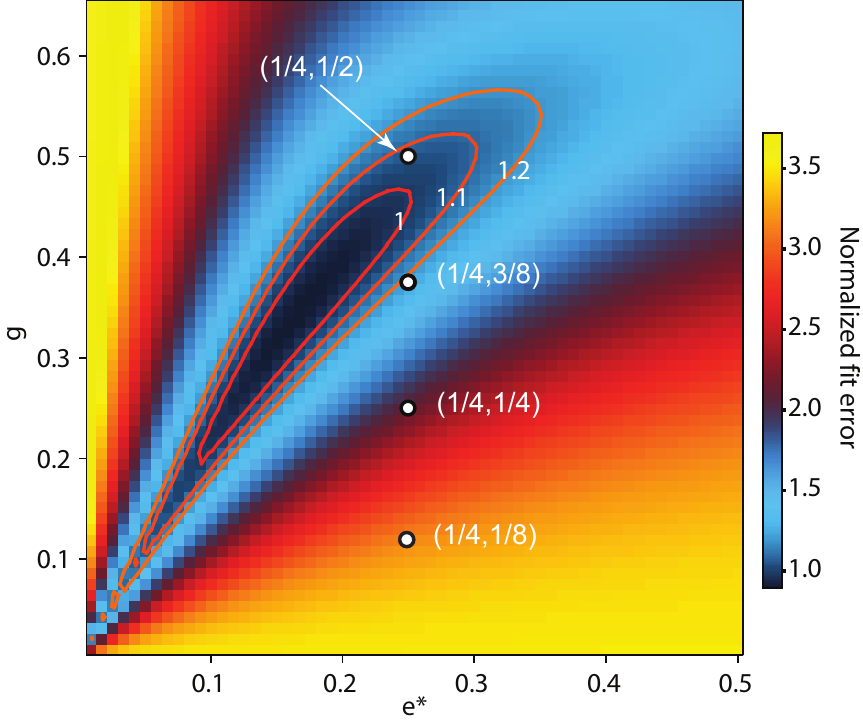}
\caption{
Map of the fit quality. Normalized fit error is the residual from the least-squares fit, divided by the number of points and by the noise of the measurement. Also marked on the map are proposed theoretical pairs ($e^*$, $g$).}
\end{figure}

Having chosen the magnetic field and gate voltage in this way, we measure the dc bias dependence in device 1 at various temperatures (Fig.~4). The traces in Fig.~4A are slices along the dashed lines in Fig.~3. Since the voltage drop between the two counter-propagating edge states in the QPC is the dc current multiplied by the Hall resistance, we have labeled the horizontal axis with both the current and the dc voltage, using $R_{xy}=0.4~$h/e$^2$\cite{wenPRB44:1991}. All these traces saturate at the same value $R^{\infty}$ at high dc bias, higher than the expected value 0.40~ h/e$^2$. The height of the peak, measured from $R^{\infty}$, decreases with increasing temperature, following a power law in temperature with exponent $-1.3$ (Fig.~4B). The full width at half maximum (FWHM) of the peak increases linearly with temperature and extrapolates to zero at zero temperature, consistent with a zero intrinsic line-width (Fig.~4C). The data can be collapsed onto a single curve (Fig.!
 ~4D) when the horizontal axis is scaled by $T$ and the vertical axis is scaled by $T^{-1.3}$ (after subtracting a common background $R^{\infty}$).

\paragraph*{Extracting g and e*.} The observed temperature dependence of the peak height and FWHM is consistent with the theoretical predictions of weak quasiparticle tunneling between fractional edge states \cite{wenPRB44:1991, bishara2, fendley2006}. In that picture, the zero-bias peak height is expected to vary with temperature as $T^{2g-2}$, which gives $g=0.35$ for the data in Fig.~4B. The weak-tunneling expression, which includes the effects of dc bias \cite{wenPRB44:1991} has the form

\begin{equation}\label{}
    g_T = A T^{(2g-2)}F(g, \frac{e^\ast I_{dc}R_{xy}}{kT}),
\end{equation}

\noindent (see Supporting Online Material for details). This functional form fits the experimental data very well, as seen in Fig.~4E (Note that $R_D$ and $g_T$ differ only by an offset and scale factor.) All five temperatures are fit simultaneously with four free parameters: a single vertical offset corresponding to $R^{\infty}$, an amplitude $A$, and the two quantities $g$ and $e^*$. A least-squares fit over the full data set gives best-fit values $g=0.35$---the same value found from the power law fit of the peak heights (Fig.~4B)---and $e^*=0.17$. Uncertainties in these values will be discussed below. Similar analysis performed on data from a different device (device 2 but energizing only gates G1 and G4) yields quantitatively similar results.

To characterize the uncertainty of these measured values, a matrix of fits to the weak-tunneling form, Eq.~(2), with $g$,  $e^*$ fixed and $A$,  $R^{\infty}$ as fit parameters is shown in Fig.~5. The color scale in Fig.~5 represents the normalized fit error, defined as the residual of the fit per point divided by 0.0005~h/e$^2$, the noise of the measurement. A fit error $\lesssim 1$ indicates that fit is consistent with the data within the noise of the measurement. Higher values indicate worse fits (see Supporting Online Material).

This matrix of fits allows various candidate states at $\nu=5/2$ to be compared with the tunneling data. All of the candidate states predict $e^*=1/4$, but $g$ can differ. States with abelian quasiparticle statistics include the (331) state \cite{Halperin_orig331,Halperin331}, which has a predicted $g=3/8$ \cite{PhysRevLett.70.355}, and the $K=8$ state with $g=1/8$\cite{K8}. States with non-abelian quasiparticle statistics include the Pfaffian \cite{moore_read} with $g=1/4$ \cite{PhysRevLett.70.355}, its particle-hole conjugate the anti-Pfaffian \cite{levin:236806,lee:236807} with $g=1/2$ \cite{levin:236806,lee:236807,bishara2}, and the $U(1)\times SU_2(2)$ state \cite{PhysRevLett.66.802}, also with $g=1/2$. Parameter pairs ($e^*$, $g$) representing these candidate states are marked in Fig.~5. Evidently, the states with $e^*=1/4$ and $g=1/2$, both non-abelian, are most consistent with our tunneling data. The abelian state with $e^*=1/4$ and $g=3/8$ cannot be excluded. We not!
 e that weak tunneling of $e^*=1/2$ quasiparticles appears inconsistent with the data, suggesting that unpaired composite fermions do not play a significant role in tunneling for this experimental situation.

\begin{figure}[b]
\includegraphics[width=3.25in]{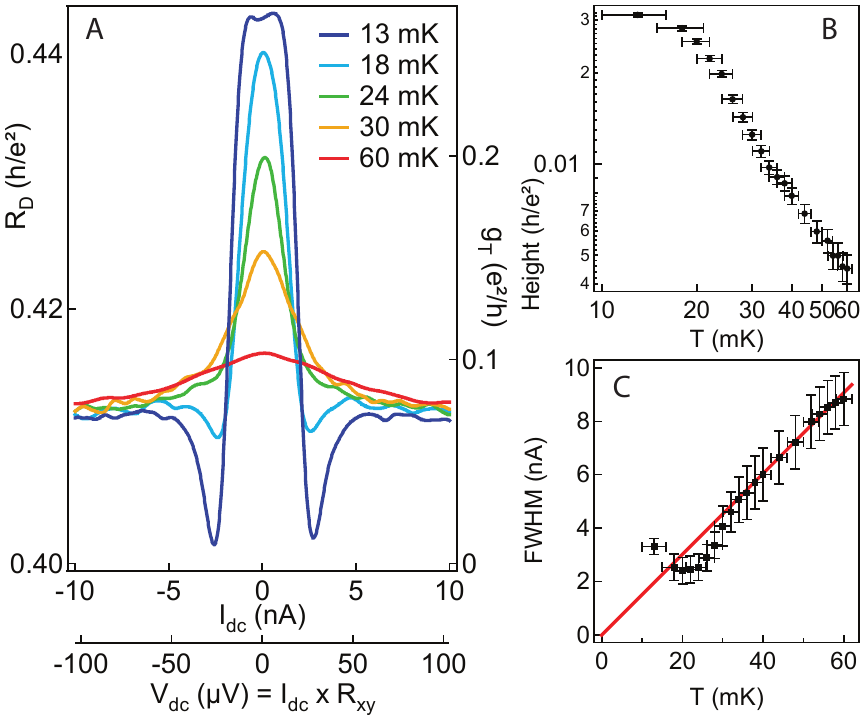}
\caption{
{\bf A.} $R_{D}$ (device 2) as a function of dc bias at fixed magnetic field ($B=4.31$~T, middle of $\nu=5/2$) and fixed gate voltage ($V_g= -2.4$~V) at several temperatures. $R_{xy}$ is independent of dc bias over this range of $I_{dc}$ (not shown). At the lowest temperature, the peak develops a flat top at a value of resistance consistent with the resistance at $\nu = 7/3$. {\bf B.} Zero-bias peak height as a function of temperature. The peak height saturates at the lowest temperatures.  {\bf C.} Peak width as a function of temperature. The red line is best fit of the high temperature data with a line going through zero. Note that below $\sim$30~mK the peak width no longer follows this line.}
\end{figure}

\paragraph*{Strong tunneling.} In contrast to device 1, the dc bias data from device 2 show evidence for strong tunneling. Device 2 has a long, channel-like geometry, which could increase the number of tunneling sites and hence the tunneling strength. Diagonal resistance, $R_D$ as a function of dc bias at several temperatures is shown in Fig.~6A. Comparing these data to those from the short QPC (Fig.~4A), shows qualitative differences at lower temperatures. At higher temperatures, the zero-bias peak height can be described by a power law in temperature with an exponent similar to that in the  QPC (Fig.~6B), and a FWHM that is proportional to temperature (Fig.~6C). At lower temperatures, the peak height deviates from a power law and saturates at the lowest temperatures at a value of resistance consistent with the resistance at $\nu=7/3$ (the resistance is higher than $3/7$ h/e$^2$ by the background $R^{\infty}-0.4$), and the FWHM deviates from the linearity seen at higher tem!
 perature. We also observe that the peak develops a flat top and strong side-dips (Fig.~6A) at the lowest temperature.

We are not aware of quantitative predictions for the strong tunneling regime for $\nu=5/2$. However, qualitative comparisons with strong tunneling theory \cite{Fendley} and experiment \cite{Roddaro1, Roddaro2, Roddaro3} at other FQH states ($\nu<1$) can be made. For strong tunneling, the edge states associated with the topmost fractional state ($\nu = 5/2$ in the present case) are backscattered almost entirely so that the quasiparticle tunneling takes place along the QPC rather than across it \cite{fendley2006, Fendley}. The flat-top peak shape and strong side dips (Fig.~6A), much stronger that expected from weak tunneling (Eq. 2), are qualitatively consistent with previous strong-tunneling studies for $\nu<1$ \cite{Roddaro2, Fendley}. The value of $R_D$ at the peak is consistent with full backscattering of the $5/2$ edge and a $\nu =7/3$ underlying edge state.

\paragraph*{Summary and outlook.}We have studied quasiparticle tunneling between edge states in a constriction at $\nu=5/2$. By annealing the device at 4K and making the constriction short we have obtained the conditions for weak tunneling, and find quantitative agreement with theory for the temperature and bias dependence of the quasiparticle tunneling peak around zero bias. From these dependences we estimate the quasiparticle charge $e^*$ and an interaction parameter $g$.  Among the candidate states, the data are most consistent with the anti-Pfaffian and $U(1)\times SU_2(2)$ both of which are non-abelian states with $e^*=1/4$ and $g=1/2$.

Beyond enabling investigations in the fundamental physics, toward a demonstration of non-abelian statistics, these experiments demonstrate a high degree of control of interedge tunneling of the $5/2$ edge state, a prerequisite for quasiparticle braiding operations needed for related schemes of topological quantum computing.

We acknowledge helpful discussions with W. Bishara, C. Chamon, C. Dillard, P. Fendley, M. Fisher, B. Halperin, E. Levenson-Falk, D. McClure, C. Nayak, B. J. Overbosch, B. Rosenow, A. Stern, X.-G. Wen, and Yiming Zhang, and experimental help from S. Amasha and A. Klust. This work was supported in part by ARO (W911NF-05-1-0062), the NSEC program of the NSF (PHY-0117795) and NSF (DMR-0353209) at MIT and by the Microsoft Corporation Project Q and the Center for Nanoscale Systems at Harvard University.

\end{document}